\documentclass[conference, letterpaper]{IEEEtran}
\IEEEoverridecommandlockouts
\usepackage{blindtext, graphicx}
\usepackage{lipsum}
\usepackage{float}
\usepackage{cuted}
\usepackage{xfrac}
\usepackage{cite}
\usepackage{amsmath,amssymb,amsfonts}
\usepackage{algorithmic}
\usepackage{graphicx}
\usepackage{textcomp}
\usepackage{etoolbox}

\BeforeBeginEnvironment{figure}{\vskip-3ex}
\AfterEndEnvironment{figure}{\vskip-4ex}

\usepackage{xcolor}
\ifCLASSOPTIONcompsoc
\usepackage[caption=false,font=normalsize,labelfon
t=sf,textfont=sf]{subfig}
\else
\usepackage[caption=false,font=footnotesize]{subfi
g}
\fi
\usepackage{indentfirst}
\def\BibTeX{{\rm B\kern-.05em{\sc i\kern-.025em b}\kern-.08em
    T\kern-.1667em\lower.7ex\hbox{E}\kern-.125emX}}

\usepackage{stfloats}

\author{\IEEEauthorblockN{
Hichem Semira}
\IEEEauthorblockA{\textit{Electronics and New Technologies Laboratory (ENT),} \\
\textit{University of Oum El Bouaghi}\\
Oum El Bouaghi, Algeria \\
email: hichem.semira@univ-oeb.dz}
\and
\IEEEauthorblockN{
Ferdi Kara}
\IEEEauthorblockA{\textit{Wireless Communication Technologies Laboratory (WCTLab),} \\
\textit{Department of Electrical and Electronics Engineering}\\
Zonguldak Bulent Ecevit University,
Zonguldak, TURKEY 67100, \\
email: f.kara@beun.edu.tr}}

\title{Error Performance of Uplink SIMO-NOMA with Joint Maximum-Likelihood and Adaptive M-PSK}

\begin{document}
\maketitle
\begin{abstract}
This paper studies the performance of the uplink Non-Orthogonal Multiple Access (NOMA) with Joint Maximum-Likelihood Detector (JML). We present a generalized upper bound of bit-error rate (BER) expression of adaptive M-ary phase-shift keying (M-PSK) over Rayleigh fading channels. Our studies are enriched through the use of a single transmitting antenna by each user and multiple receiving antennas at the base station (single-input-multiple-output (SIMO)). The derivation of the upper bound expression is obtained considering an arbitrary constellation size by performing Maximum Ratio Combining (MRC) of diversity paths. The extensive computer simulations validate the analysis and it is revealed that the JML outperforms the existing detecting algorithms significantly and achieves the full diversity order for each user.
\end{abstract}
\begin{IEEEkeywords}
Non-Orthogonal Multiple Access (NOMA), adaptive M-PSK, multi-user detection, SIMO, Bit error rate
\end{IEEEkeywords}
\section{Introduction}
Recently the rapid growth demand of data mobile and internet has prompted the networks and communications community to find new technologies. Achieving ultra-low latency and high reliability communication in a massive connection between users and machines are among these concerns. To achieve these goals, Non-Orthogonal Multiple Access (NOMA) is strongly proposed. Currently, NOMA is receiving immense attention in academia due to its superior spectral efficiency and improvement of system throughput \cite{b1}. In NOMA, all users can share the entire frequency and time resources by allocating different power levels or multiplexing their signals in code domain \cite{b2}. Typically, the NOMA allocates more power on the worst channel conditions to the delicate user. At the receiver side to decode the superposed information, the Successive Interference Cancellation (SIC) technique, has been adopted to eliminate the interferences generated due to superposition of data \cite{b3}. In the first step, the signal with higher signal-to-interference plus noise ratio (SINR) is decoded by treating other signals as noise. In the second step, an iterative SIC technique is implemented by subtracting the previously-detected signals to decode remaining users' signal.

In the literature for NOMA systems, the performances are evaluated largely in terms of outage probability and capacity with perfect and imperfect channel state information (CSI) e.g. \cite{b4,b5,b6}, and the superiority of NOMA to related OMA (Orthogonal multiple access) systems is proved. Closed-form expressions have been obtained for the outage probability by considering independent identically and not identically distributed fading environment.

On the other hand, rarely papers like \cite{b7,b8,b9,b10} have derived an analytical expression for Bit Error Rate (BER) over fading channels. These studies have given an asymptotic and a closed-form BER expressions for NOMA systems with SIC \cite{b7,b8,b10}. However, the SIC-based detection has a poor error performance in the uplink path and suffers from the error floor in high SNR \cite{b7,b8} although it has shown its effectiveness in downlink NOMA. Recently to overcome this problem in the uplink NOMA, the Joint Maximum Likelihood (JML) detector is proposed instead of SIC detector \cite{b9,b10,b11} and its superiority has been proved. 
In \cite{b9}, an upper bound analysis of JML detector is investigated. However, this analysis is devoted only for the QPSK modulation and its performance on the higher modulation is not well-described. Nevertheless, in the standards \cite{b12}, to enhance the spectrum efficiency, an adaptive modulation is considered instead of fixed modulations order, where based on CSI estimated by the base station, each user can increase or decrease the rate of transmission by adapting an appropriate modulation and coding rate scheme to the quality of the wireless link. To this end, in this paper, we describe a performance study of the JML technique assuming a higher order M-ary Phase Shift Keying (M-PSK) modulation with a size $ M\geq 4$. And to make the study more comprehensive for standards, we perform analysis based on adaptive modulation.
We derive a general upper bound BER expression by considering a Maximum Ratio Combining for multiple antennas and a JML (MRC-JML) detector at the receiver with an adaptive M-PSK modulation, where users send their data using different constellation sizes mapping with Gray code.

The remainder of this paper is organized as follows. In Section II, the system of uplink single-input-multiple-output (SIMO)-NOMA is presented with joint JML and the signal models: noise and channel. In Section III, the derivation of upper bound of the BER by using union bound is discussed. In Section IV computer simulation results are presented and compared with analytical results for different constellation sizes. Finally, the conclusion is drawn in Section V.

\begin{figure}[t]
\includegraphics[width=8cm]  {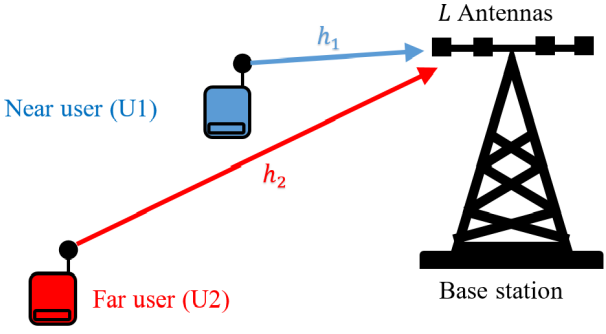}
\caption{Illustration of uplink SIMO-NOMA} 
\label{fig1}
\end{figure}
\section{System Model}
In this paper, we consider a multi-access transmission, corresponding to uplink NOMA scenario, where two users equipped with a single antenna to transmit their signals to a base station equipped with $L$ antennas (SIMO-NOMA) (see Fig. 1). The two users sharing the same uplink channel and can transmit with their own powers on the same frequency block within the same time slot. For simplicity, we assume that both users send their messages with the same power $P_1=P_2=P$. The received signal by the base station $\textbf{y}\in\mathbb{C}^{L\times1}$ is given by
\begin{equation}\label{eq:1}
\mathbf{y}=\sqrt{P} \mathbf{h}_1 x_1 + \sqrt{P} \mathbf{h}_2 x_2+\mathbf{w},
\end{equation}
where $\mathbf{h}_i=[h_{i1}\cdots h_{iL} ]^T$  denotes the multipath fading channel vector between $i^{th}$ user and BS. The symbol $[\;]^T$ defines the transpose of a vector. The coefficients $h_{il},l=1,\cdots,L$ are independent and identically distributed (i.i.d) and  modeled as complex Gaussian random variables with zero mean and variance $\sigma_i^2$, i.e., $\mathbf{h}_i \sim  \mathcal{CN}(0,\sigma_i^2\mathbf{I}_L)$, where  $\mathbf{I}_L$ denotes the $L \times L$ identity matrix. We note that $\sigma_i^2=\mu d_i^{-\lambda}$ where $d_i$ is the distance between $i^{th}$ user and BS and $\mu$ and $\lambda$ are called the propagation constant and the path loss exponent, respectively. In uplink-NOMA scenario it is assumed that  $d_1<d_2$. Therefore, the coefficients of the channel verify the relation $\left\|\mathbf{h}_1\right\|>\left\|\mathbf{h}_2\right\|$, where the symbol $\left\|.\right\|$ denotes the Frobenius norm.
Hence, the user having the large channel gain is called Near user or intra-cell user  (user1-U1) and the one exposed to the small channel gain is called Far user or cell-edge user (user2-U2). The additive white Gaussian noise (AWGN) vector is denoted by $\mathbf{w}=[w_1\cdots w_L ]^T$ where each component  is modeled as complex Gaussian random variable with $w_l \sim  \mathcal{CN}(0,\frac{N_0}{2})$.

The information is modulated as complex symbols $x_i\in \mathbb{C}, i=1,2$, from constellation of size $M_i,i=1,2$, where $M_i$  is a power of $2$.  In this paper, we consider an adaptive modulation that uses $M_i$-PSK with Gray code mapping. Thus, the symbols $x_i$ take their value from the alphabet $\chi_i=\{s_{i1},s_{i2},\cdots,s_{iM_i}\}$, where
\begin{equation} \label{eq:2}
\begin{split}
s_{im_{i}}=\exp{\left(j(m_i-1)\frac{2\pi}{M_i}+j\frac{\pi}{M_i}\right)},\qquad{} &m_i=1,\cdots,M_i.\\&i=1,2.
\end{split}
\end{equation}

It is possible that both users take their symbols from the same constellation alphabet, i.e., $M_1=M_2$, or may take their symbols from different constellation alphabet, i.e., $M_1 \ne M_2$. In the last case, since the U2 exhibits worse channel condition than U1, we consider $M_1>M_2$. In order to detect the superposed information at the BS, we can use one of the following detector
\subsection{MRC-SIC detector}
SIC is based on the exploitation of the properties of the differences in signal power. Assuming that the base station knows the channels coefficients $\mathbf{h_i}$, firstly, the BS tries to detect U1 symbols via a minimum-distance criterion \cite{b13} by considering the signal of U2 as noise
\begin{equation} \label{eq:3}
\Hat{x}_1 = \underset{x_1\in \chi_1}{\mathrm{argmin}}\left\|\textbf{y}-\textbf{h}_1 x_1\right\|^2,
\end{equation}
where $\chi_1$ is the set of all  possible symbols from a constellation points of size $M_1$. Secondly, the BS tries to detect U2 symbols after having subtracted the signal of U1 from $\mathbf{y}$ by using the same procedure
\begin{equation}\label{eq:4}
\Hat{x}_2 = \underset{x_2\in \chi_2}{\mathrm{argmin}}\left\|\left(\textbf{y}-\textbf{h}_1 \Hat{x}_1\right)- \textbf{h}_2 x_2\right\|^2,
\end{equation}
where $\chi_2$ is the set of all the possible symbols from constellation of size $M_2$.
\subsection{MRC-JML detector}
Like the MRC-SIC, assuming that the channel coefficients $\mathbf{h}_i$ are known at BS, the JML detector performs an exhaustive search to simultaneously decode the emitted signals $x_1$ and $x_2$ as fallows
\begin{equation} \label{eq:5}
\lbrack\Hat{x}_1,\Hat{x}_2\rbrack= \underset{x_1\in \chi_1,x_2\in \chi_2}{\mathrm{argmin}}\left\|\textbf{y}-\textbf{h}_1 x_1-\textbf{h}_2 x_2\right\|^2.
\end{equation}

The JML carries out an exhaustive joint search for all the possible combinations of the superposition of the two vectors $\lbrack\mathbf{h}_1x_1,\mathbf{h}_2x_2\rbrack$, that makes it immune from the error floor, unlike the SIC detector, which suffers from the propagation of the error floor in high SNR.
\section{Error Probability Analysis}
In this section, we derive an upper bound of the BER by considering an adaptive M-ary PSK Modulation. We first define the upper bound for the U1, and then we deduce that of the U2. We illustrate the different error patterns of superposed modulations based on the characteristics of Gray code. From these regularities, we formulate the upper bound expression.
\subsection{Upper bound of BER for Near user}
For the illustration simplicity, let us consider an 8-PSK for U1  $(M_1=8)$ and an 4-PSK for U2  $(M_2=4)$. The Fig. 2.a shows the superposed symbols from both users at the BS, where the first three bits belong to U1 and the last two bits belong to U2. The binary bit representations of the two symbols are given in the form of ${\{b_{11} b_{12} b_{13} b_{21} b_{22}\}}$, where the first sub-index represents the user and the second represents the bit arrangement inside the received symbol.
\begin{figure*}
\centering
\subfloat[{Error pattern for Near user (U1)}]{\includegraphics[width=6cm]{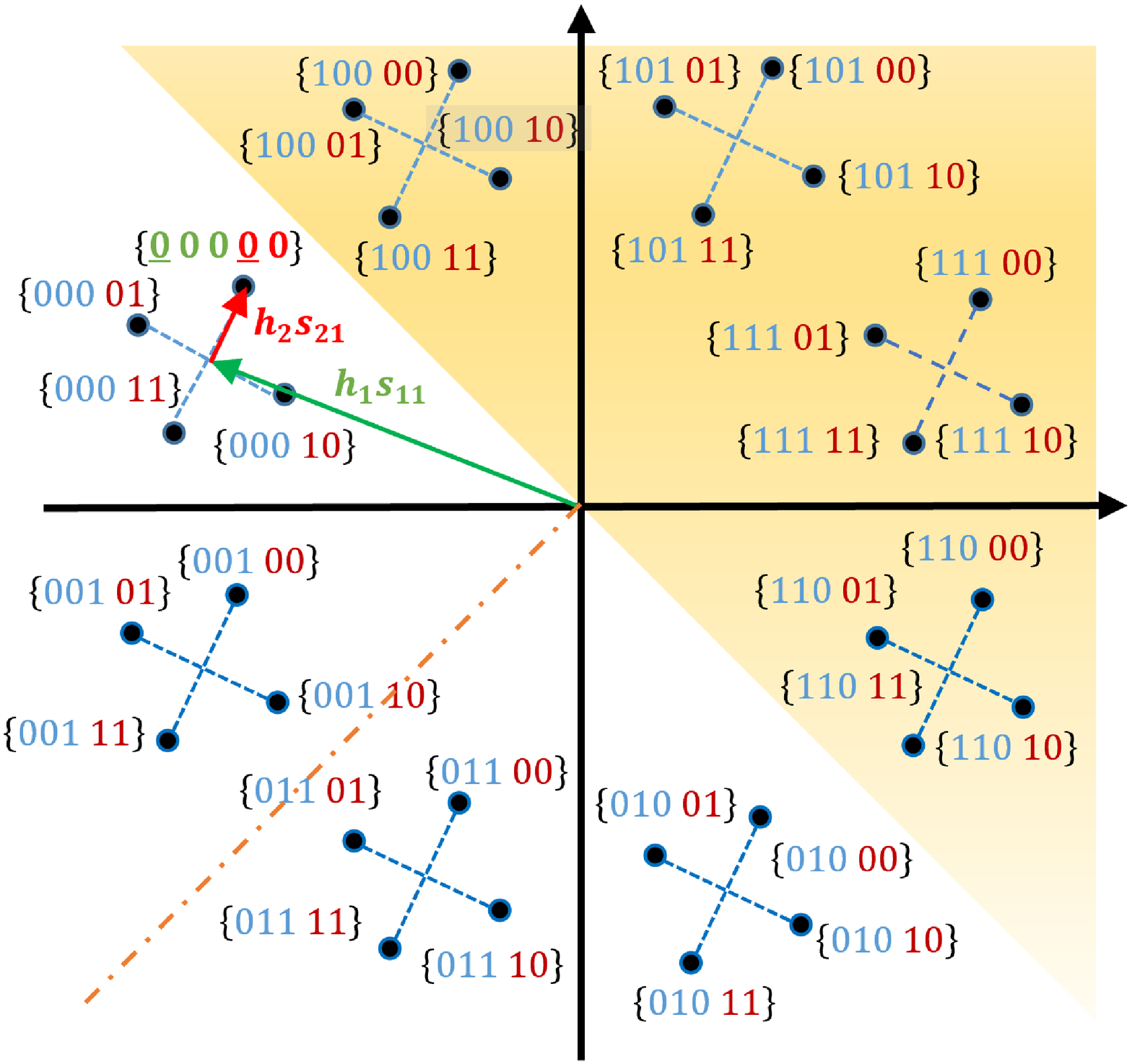}
\label{const1}}
\subfloat[{The effective distances for Near user (U1)}]{\includegraphics[width=6cm]{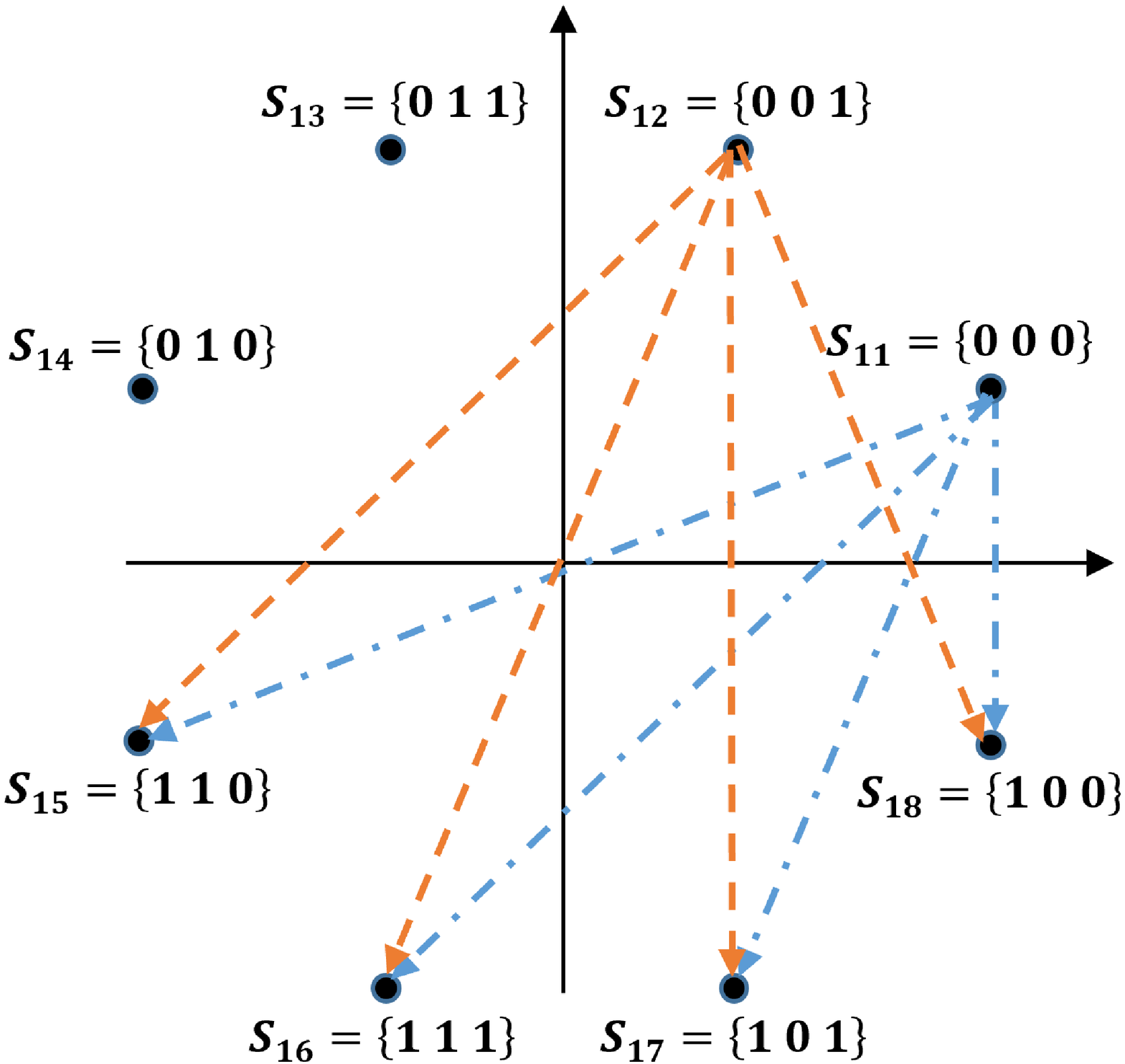}
\label{const2}}
\subfloat[{Error pattern for Far user (U2)}]{\includegraphics[width=6cm]{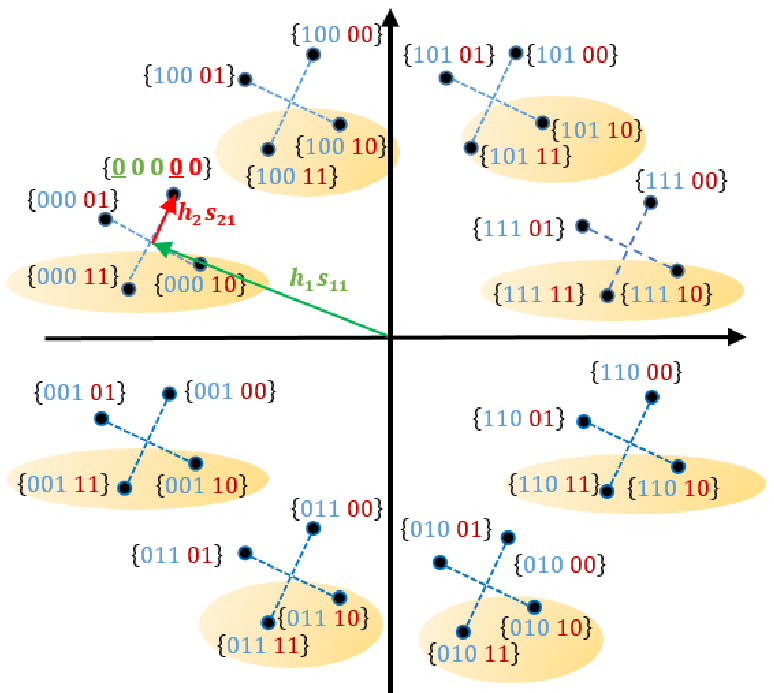}
\label{const3}}
\caption{{Signal space diagram for superposed 8-PSK symbols with 4-PSK symbols based on Gray code.}}
\label{constellations}
\end{figure*}

To make things simpler and without loss of generality, we suppose that both users send all zero bits with unity power by following their mapping rule, $x_1= s_{11}=\exp\left(j \frac{\pi}{8} \right)$ for U1 and $x_2=s_{21}= \exp\left(j \frac{\pi}{4} \right)$ for U2. Let’s now concentrate our BER analysis on the first bit $\left(b_{11}=0\right)$ of the symbol $x_1$. As the BS makes a decision according to ML detector, it is clear that an error occurs when the JML selects one of the symbols found in the shaded area of Fig. 2.a, which are characterized  by their first bit $\left(b_{11}=1\right)$, i.e., the lower half-circle of an 8-PSK constellation with Gray code (the four symbols with first bit equal to one). It is straightforward to notice that the total number of erroneous symbols after making a wrong estimation is equal to the all U2 symbols superimposed with the four symbols of U1  ($\frac{8}{2}\times4$ symbols). By the same manner, we can show that the number of errors for any $M_1$-ary PSK Modulation superimposed with $M_2$-ary PSK Modulation is equal to $\frac{M_1M_2}{2}$ .

The pairwise error probability (PEP) of making wrong decision between $s_{11}$ and $s_{1m_1}$ conditioned on the channels gain $\textbf{h}_1$ and $\textbf{h}_2$ \cite{b14} is given by
\begin{equation} \label{Eq6}
\begin{split}
    &Pr\left({s_{11}\to s_{1m_1}} \mid {\mathbf{h}_1,\mathbf{h}_2}\right)=\\&\qquad {}Q\left(\frac{\parallel {\mathbf{h}}_1\left(s_{11}-s_{1m_1}\right)+\mathbf{h}_2\left(s_{21}-s_{2m_2}\right)\parallel
}{\sqrt{2  N_0}}\right).\\
&m_1=\frac{M_1}{2}+1,\cdots, M_1. \qquad{}  m_2=1,\cdots, M_2.
\end{split}
\end{equation}

We note that we have in total $\frac{M_1\times M_2}{2}$ conditioned PEP. Let’s now denote the distances between the symbol $s_{11}$ and symbols ${s_{1m_1}}$, $m_1=\frac{M_1}{2}+1,\cdots, M_1$, by $\mathfrak{d}_k^1=\| \mathbf{h}_1 \mathfrak{a}_k^1+\mathbf{h}_2 \mathfrak{b}_k \|$, $k=1,\cdots, \frac{M_1M_2}{2}$ where  $\mathfrak{a}_k^1=\left({s_{11}}-{s_{1_{\frac{M_1}{2}+n}}}\right)$, for  $(n-1) M_2+1\leq k \leq n M_2$, $n=1,\cdots, \frac{M_1}{2}$, and $\mathfrak{b}_k=\left ({s_{21}}-{s_{2m_2}} \right)$, $m_2=1,\cdots, M_2$, within each sub-interval of order $n$. It is worth noting that for a given $\mathbf{h}_1$ and $\mathbf{h}_2$, the distances $\mathfrak{d}_k^1$ depend only on $\mathfrak{a}_k^1$ and  $\mathfrak{b}_k$ which are computed from the alphabet $\chi_1$ and $\chi_2$, thus, $\mathfrak{d}_k^1$ are the distances between $M_i$-PSK symbols in a new space rotated by $\mathbf{h}_1$ and $\mathbf{h}_2$, and as all the symbols have undergone the same rotation (a new space after rotation by $\mathbf{h}_1$ for U1 and $\mathbf{h}_2$ for U2), hence, the constellation points keep the same symmetry property and error patterns.

Until now, we have derived the conditional PEP when $x_1$ and $x_2$ are equal to $s_{11}$ and $s_{21}$, respectively. If we consider the transmission of $s_{11}$ superposed with any other symbols ${s_{2m_2}}$, $m_2=2,\cdots, M_2$, and because of the circularity and  symmetry of $M_2$-PSK constellation, we obtain the same values for $\mathfrak{b}_k$ with different order within each $\frac{M_1}{2}$ sub-intervals of size $M_2$. In all cases the conditional PEP is the same without considering the order.

Concerning the remaining superimposed symbols ${s_{1m_1}}$, $m_1=2,\cdots, \frac{M_1}{2}$, regardless of ${s_{2m_2}}$, $m_2=1,\cdots, M_2$, from the $M_2$-PSK  constellation, the number of errors is always the same and equal $\frac{M_1\times M_2}{2}$.
Furthermore, the conditional PEP $Pr\left({s_{1m_1}}\to {s_{1_{\frac{M_1}{2}+n}}} \mid {\mathbf{h}_1,\mathbf{h}_2}\right)$, $m_1=1,\cdots, \frac{M_1}{2}$, $n=1,\cdots, \frac{M_1}{2}$,  is uniquely different according to values of $\mathfrak{a}_k^{m_1}$, $m_1=1,\cdots, \frac{M_1}{2}$, as long as the values of $\mathfrak{b}_k$  are always the same. It is easy to show again from $M_1$-PSK constellation that there are a total of $\frac{M_1}{4}$ different values of $\mathfrak{a}_k^{m_1}$, $k=1,\cdots, \frac{M_1M_2}{2}$, $m_1=1,\cdots, \frac{M_1}{4}$, because of symmetry, the  distances between the points which are located in upper half-circle (first bit $b_{11}=0$) and those located in lower half-circle (first bit $b_{11}=1$)  are equal in pairs with respect to the right and left semicircle (see Fig. 2.b), accordingly, for the upper bound computation we only consider the contribution of $\frac{M_1}{4}$ symbols located in the right quadrant.

At this point, we have considered the case where the first bit was a zero, in the case where it was a one, owing to the symmetry we find the same conditional PEP. Therefore, we have only discussed the errors corresponding to first bit $b_{11}$. Furthermore, if we consider the rest of bits  $b_{1_{\log_2{M_1}}}$ and  by exploiting the symmetry of $M_1$-PSK constellation,  \cite[Eq. (10)]{b15}  has found an approximated expression of the probability of error depends only on the right quadrant, i.e., exploiting only the $\frac{M_1}{4}$ symbols to drive the distances to the points located in lower half-circle. Assuming equally likely symbols and take on consideration the new space diagram, we have

\begin{equation} \label{eq:7}
Pr\left(e\mid {\mathbf{h}_1,\mathbf{h}_2}\right) \cong \frac{M_1}{2\log_2{M_1}}P_a,
\end{equation}
where
\begin{equation} \label{eq:8}
\begin{split}
P_a=\frac{4}{M_1}\sum_{{m_1}=1}^{\frac{M_1}{4}}Pr\left({s_{1m_1}}\to {s_{1_{\frac{M_1}{2}+n}}}\mid {\mathbf{h}_1,\mathbf{h}_2}\right),\\ \qquad{} n=1,\cdots, \frac{M_1}{2},
\end{split}
\end{equation}
or
\begin{equation} \label{eq:9}
P_a=\frac{4}{M_1}\sum_{{m_1}=1}^{\frac{M_1}{4}}Q\left(\frac{\mathfrak{d}_k^{m_1}}{\sqrt{2N_0}}\right),\quad k=1,\cdots, \frac{M_1M_2}{2}.
 \end{equation}
By combining (\ref{eq:7}) and (\ref{eq:9}), we get
\begin{equation}\label{eq:10}
\begin{split}
  Pr\left(e\mid {\mathbf{h}_1,\mathbf{h}_2}\right) \cong \frac{2}{\log_2{M_1}}\sum_{{m_1}=1}^{\frac{M_1}{4}}Q\left(\frac{\mathfrak{d}_k^{m_1}}{\sqrt{2N_0}}\right),\\\qquad{}
  k=1,\cdots, \frac{M_1M_2}{2}.
  \end{split}
\end{equation}

Using the upper bound to the probability of a union of events, it can be written as
\begin{equation}\label{eq:11}
 Pr\left(e\mid {\mathbf{h}_1,\mathbf{h}_2}\right) \leq \frac{2}{\log_2{M_1}}\sum_{{m_1}=1}^{\frac{M_1}{4}} \sum_{k=1}^{\frac{M_1M_2}{2}}Q\left(\frac{\mathfrak{d}_k^{m_1}}{\sqrt{2N_0}}\right),
\end{equation}
where $\mathfrak{d}_k^{m_1}=\| \mathbf{h}_1 \mathfrak{a}_k^{m_1}+\mathbf{h}_2 \mathfrak{b}_k \|$, $k=1,\cdots, \frac{M_1M_2}{2}$, $\mathfrak{a}_k^{m_1}=\left({s_{1m_1}}-{s_{1{\frac{M_1}{2}+n}}}\right)$, for $(n-1) M_2+1\leq k \leq n M_2$, $n=1,\cdots, \frac{M_1}{2}$, and $\mathfrak{b}_k=\left ({s_{21}}-{s_{2m_2}} \right )$, $m_2=1,\cdots, M_2$, remain unchanged within each sub-interval of order $n$. As the complex vector $\mathbf{h}_1$, $\mathbf{h}_2$ are modeled as complex Gaussian random vector, thus $\mathbf{h}_1 \mathfrak{a}_k^{m_1}+\mathbf{h}_2 \mathfrak{b}_k \sim  \mathcal{CN}(0,\sigma_1^2 |\mathfrak{a}_k^{m_1}|^2+\sigma_2^2 |\mathfrak{b}_k|^2)$, where $|\mathfrak{a}_k^{m_1}|=2|\sin\left({\left(m_1-\frac{M_1}{2}-n\right)\frac{\pi}{M_1} }\right)|$ and $|\mathfrak{b}_k|=2|\sin\left({\left(1-{m_2}\right)\frac{\pi}{M_2} }\right)|$, $m_2=1,\cdots, M_2$.

If we write \cite{b9} $Z_k=\frac{\left({\mathfrak{d}_k^{m_1}}\right)^2}{2 N_0} $ and consider $L$ uncorrelated signals $\textbf{y}\in\mathbb{C}^{L\times1}$ received at BS, then $Z_k$ obeys Erlang distribution \cite[Eq. (10.61)]{b16} with  the PDF
\begin{equation} \label{eq:12}
 P_{Z_k}\left(z\right)=\frac{1}{\left(L-1\right)!}\frac{z^{L-1}}{{\Gamma^{L}_{{m_1}k}}} \exp{\left(\frac{-z}{\Gamma_{m_1k}}\right)},
 \end{equation}
where $\Gamma_{{m_1}k}=\frac{\sigma_1^2 |\mathfrak{a}_k^{m_1}|^2+\sigma_2^2 |\mathfrak{b}_k|^2}{2}$. The upper bound is computed by averaging the BER over the distribution of $Z_k$ as
\begin{equation} \label{eq:13}
  P_1(e)\leq \frac{2}{\log_2{M_1}}\sum_{{m_1}=1}^{\frac{M_1}{4}} \sum_{k=1}^{\frac{M_1M_2}{2}}\int_{0}^\infty Q\left(\sqrt{z}\right)
 P_{Z_k}\left(z\right) dz,
\end{equation}
Following \cite{b9}, we get the (\ref{eq:14}) (see the top of the next page).

It can be seen from Fig.  2.a and Fig.  2.b that the distances
$\left({s_{11}}\to {s_{1m_1}}\right)$, $m_1= \frac{M_1}{2}+1,\cdots, M_1$, are the smallest distances compared to other distances that are in the same quadrant, consequently, their  PEP $Pr\left({s_{11}}\to s_{1m_1}\mid {\mathbf{h}_1,\mathbf{h}_2}\right)$ dominate the BER for high SNR, then The upper bound (\ref{eq:14}) can be roughly replaced by (\ref{eq:15}) (see the top of the next page).

The expressions (\ref{eq:14}) or (\ref{eq:15}) are the general expressions of BER upper bound for U1 utilizing $M_1$-ary PSK, and it is clear that for $M_1=4$, the expressions (\ref{eq:14}) or (\ref{eq:15}) are reduced to the upper bound found in \cite[Eq. (7)]{b9} which proves our analysis.

\begin{figure*}
    \begin{equation} \label{eq:14}
 P_1(e)\leq\frac{1}{\log_2{M_1}}\sum_{{m_1}=1}^{\frac{M_1}{4}} \sum_{k=1}^{\frac{M_1M_2}{2}} \left[ 1-\sum_{l=0}^{L-1}{2l\choose l} \sqrt{\frac{1}{1+2/\Gamma_{m_1k}}} \frac{1}{\left(2\Gamma_{m_1k}+4\right)^l} \right],\quad M_1\geq 4. \qquad{}
\end{equation}
\end{figure*}
\begin{figure*}
\begin{equation}\label{eq:15}
    P_1(e)\leq\frac{1}{\log_2{M_1}} \sum_{k=1}^{\frac{M_1M_2}{2}} \left[ 1-\sum_{l=0}^{L-1}{2l\choose l} \sqrt{\frac{1}{1+2/\Gamma_{1k}}} \frac{1}{\left(2\Gamma_{1k}+4\right)^l} \right],\quad M_1\geq 4. \qquad{}
\end{equation}
\end{figure*}
\subsection{Upper bound of BER for Far user}
Now we refer to Fig. 2.c, and consider that both users transmit all zero bits. As in previous subsection, assume that U1 chooses its symbols from an octophase signals and U2 from a quadrature phase signals. Like the above analysis, we first begin our study on the first bit $b_{21}$. An error of detection occurs if the JML detector chooses one of the symbols in the colorful ellipse of the Fig. 2.c, and the number of erroneous detection is $8 \times\frac{4}{2}$. Thus, if we suppose a superposition of $M_1$-ary PSK Modulation with $M_2$-ary PSK Modulation the number of errors is $\frac{M_1\times M_2}{2}$. Consequently, the conditioned PEP of confusing the symbol  $s_{21}$ with the symbols $s_{2m_2}$, $m_2=\frac{M_2}{2}+1,\cdots, M_2$, becomes for U2 as
\begin{equation} \label{eq:16}
\begin{split}
 Pr\left(s_{21}\to s_{2m_2} \mid {\mathbf{h}_1,\mathbf{h}_2}\right)=Q\left(\frac{\delta_k^1}{\sqrt{2  N_0}}\right),\\\qquad{}
  k=1,\cdots, \frac{M_2M_1}{2},
  \end{split}
\end{equation}
where $\delta_k^{1}=\| \mathbf{h}_1 \alpha_k + \mathbf{h}_2 \beta_k^{1} \|$, and $\alpha_k=\left(s_{11}-s_{1n}\right)$, for $(n-1)  \frac{M_2}{2}+1\leq k \leq n \frac{M_2}{2}$, $n=1,\cdots, M_1$,  and $\beta_k^1=\left(s_{21}-s_{2{\frac{M_2}{2}}+\mu}\right)$,  $\mu=1,\cdots, \frac{M_2}{2}$, within sub-intervals of order $n$ and of size $\frac{M_2}{2}$. The BER upper bound  of U2 can be derived via steps similar to those discussed in the analysis of U1. Therefore, under assumption that $\mathbf{h}_i \sim  \mathcal{CN}(0,\sigma_i^2\mathbf{I}_L)$, $i=1,2$, and by a full scan of all points of the $M_2$-PSK constellation with all bits $b_{2{\log_2{M_2}}}$ tested, the formula (\ref{eq:14}) for equally likely symbols becomes  as (\ref{eq:17}) (see the top of the next page),
\begin{figure*}
    \begin{equation} \label{eq:17}
 P_2(e)\leq\frac{1}{\log_2{M_2}}\sum_{{m_2}=1}^{\frac{M_2}{4}} \sum_{k=1}^{\frac{M_1M_2}{2}} \left[ 1-\sum_{l=0}^{L-1}{2l\choose l} \sqrt{\frac{1}{1+2/\overline{\Gamma}_{{m_2}k}}} \frac{1}{\left(2\overline{\Gamma}_{m_2k}+4\right)^l} \right],\quad M_2\geq 4. \qquad{}
\end{equation}
\end{figure*}
where $\overline{\Gamma}_{m_2k}=\frac{\sigma_1^2 |\alpha_k|^2+\sigma_2^2 |\beta_k^{m_2}|^2}{2}$, and
$|\alpha_k|=2\lvert\sin\left({\left(1-n\right)\frac{\pi}{M_1} }\right)\rvert$, for $(n-1)\frac{M_2}{2}+1\leq k\leq n \frac{M_2}{2}$, $n=1,\cdots, M_1$, and $\lvert\beta_k^{m_2}\rvert=2\lvert \sin \left({\left({m_2}-\frac{M_2}{2}-\mu\right)\frac{\pi}{M_2} }\right)\rvert$, $\mu=1,\cdots,\frac{M_2}{2}$, within sub-intervals of order $n$.
Like the formula (\ref{eq:15}) if we take into account only the dominant PEP terms belonging to the received signals $\mathbf{h}_1 s_{11}+\mathbf{h}_2 s_{21}$, and for high SNR the formula (\ref{eq:15}) is replaced by (\ref{eq:18}) (see the top of the next page).
\begin{figure*}
\begin{equation}\label{eq:18}
    P_2(e)\leq\frac{1}{\log_2{M_2}} \sum_{k=1}^{\frac{M_1M_2}{2}} \left[ 1-\sum_{l=0}^{L-1}{2l\choose l} \sqrt{\frac{1}{1+2/\overline{\Gamma}_{1k}}} \frac{1}{\left(2\overline{\Gamma}_{1k}+4\right)^l} \right],\quad M_2\geq 4. \qquad{}
\end{equation}
\hrulefill
\end{figure*}

As noted before, the expressions (\ref{eq:17}) or (\ref{eq:18}) are the general expressions of the BER upper bound for U2 using $M_2$-ary PSK, and if we let $M_2=4$ the expressions (\ref{eq:17}) or (\ref{eq:18}) are equal to the upper bound computed in \cite[Eq. (10)]{b9}.
\section{Numerical Results}
To validate our analysis, computer simulations were conducted with $L=4$ antennas at BS. There are
two users transmitting their signals to the BS with equal power by employing an adaptive modulation technique. To compare the obtained theoretical results with that of the simulation, we assume during simulation that the SNR is normalized and vary  the dominant gain channel  $\sigma_1^2$  for U1 taking into consideration that $\sigma_2^2=\sigma_1^2/8$ for U2. The validation of the theoretical results are carried out by considering two scenarios, one with $M= M_1=M_2$ and the other with $M_1>M_2$.

For the first scenario, the Fig. 3 shows the BER performance of the MRC-JML detector with that of MRC-SIC detector for $M=16$, $M=32$ and $M=64$, respectively. It is clear from the figures that our calculated BER upper bound using expressions (\ref{eq:14})-(\ref{eq:15}) and (\ref{eq:17})-(\ref{eq:18}) coincide well with the results obtained in the simulation. The upper bounds are very tight in the high SNR regime and it is also common for all upper bound analysis in the all communication systems. We also can see that the performance of the near user surpasses that of the far user due to the channel quality differences. If we only consider the distances which are affected by the dominant gain of channel $\sigma_1^2$ to compute the PEP, this out-performance of U1 can be explained by the use $\frac {M}{2}$ $|\mathfrak{a}_k^{m_1}|$ distances, instead of $M$ $|\alpha_k |$ distances used by the U2 with the same M-ary PSK constellation. The figures also provide a comparison between JML and SIC, and show that the MRC-JML detector outperforms the MRC-SIC significantly. The error floor problem of MRC-SIC in the high SNR regime is completely eliminated by the MRC-JML. Besides, one can easily see that the MRC-JML achieves the full diversity order (i.e., $L$) for both users. The full diversity order is observed by examining the slope of the BER curves in figures.
\begin{figure*}[htbp]
\centering
\subfloat[{$M=M_1=M_2=16$}]{\includegraphics[width=0.33\textwidth]{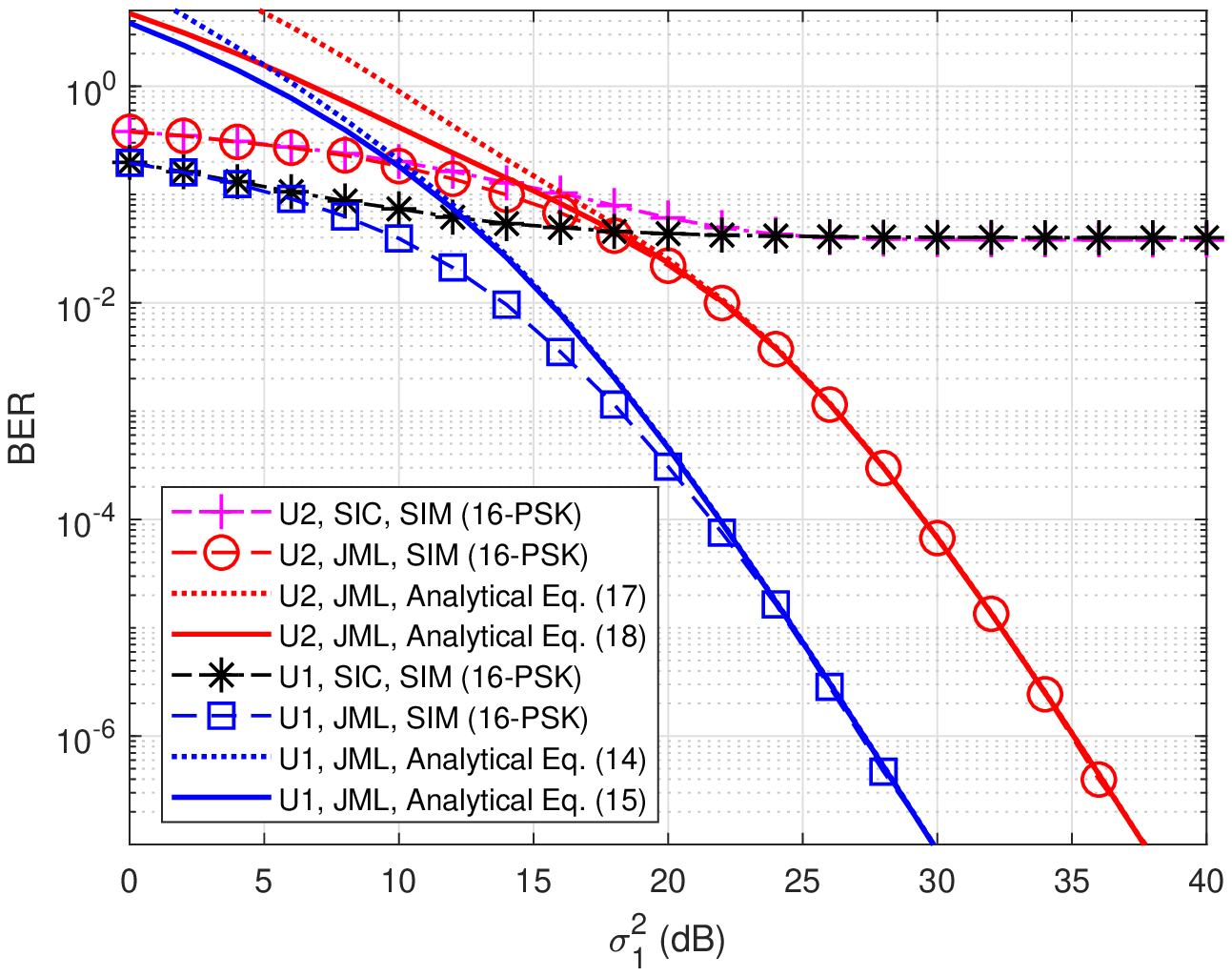}
\label{ber1}}
\subfloat[{$M=M_1=M_2=32$}]{\includegraphics[width=0.33\textwidth]{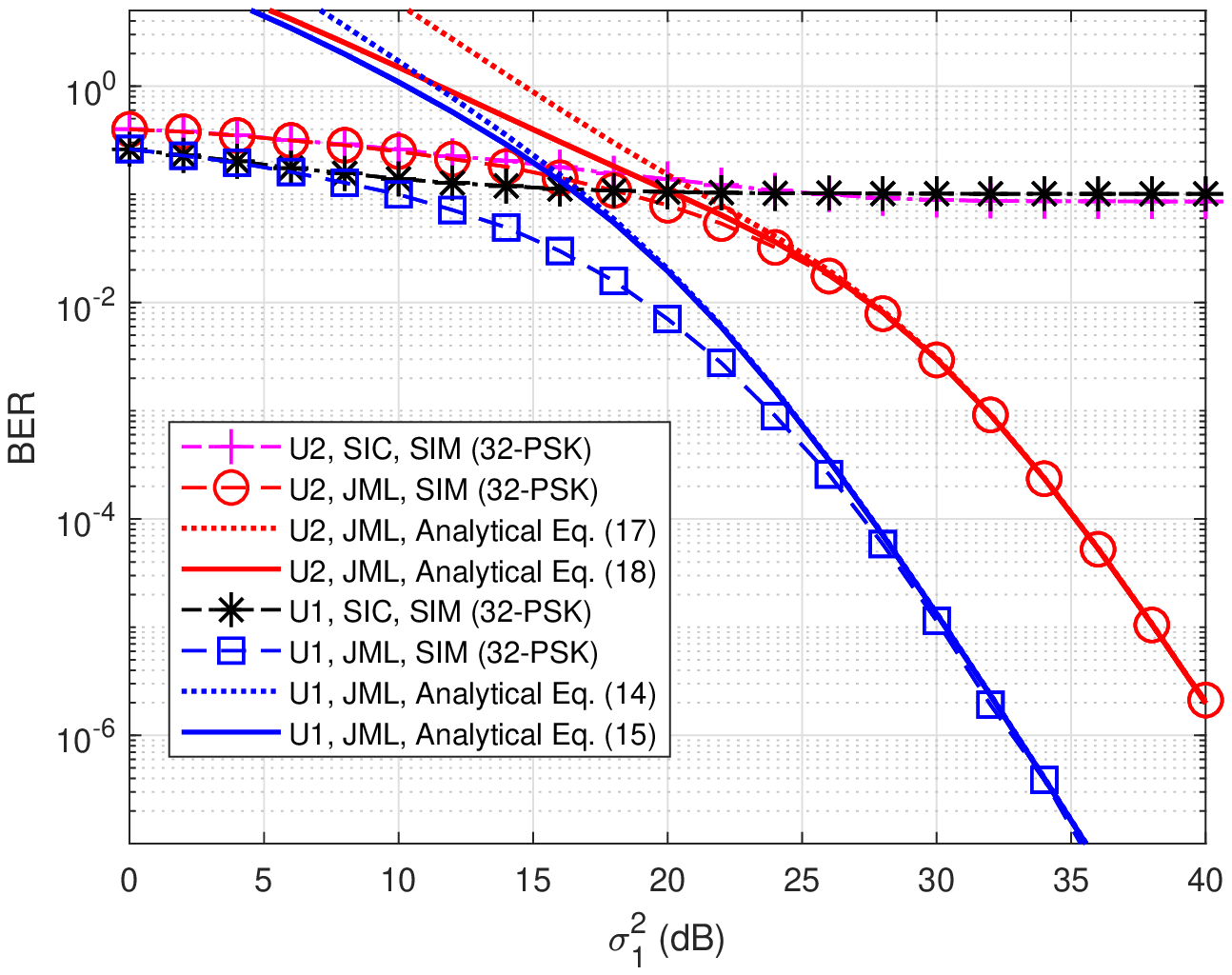}
\label{ber2}}
\subfloat[{$M=M_1=M_2=64$}]{\includegraphics[width=0.33\textwidth]{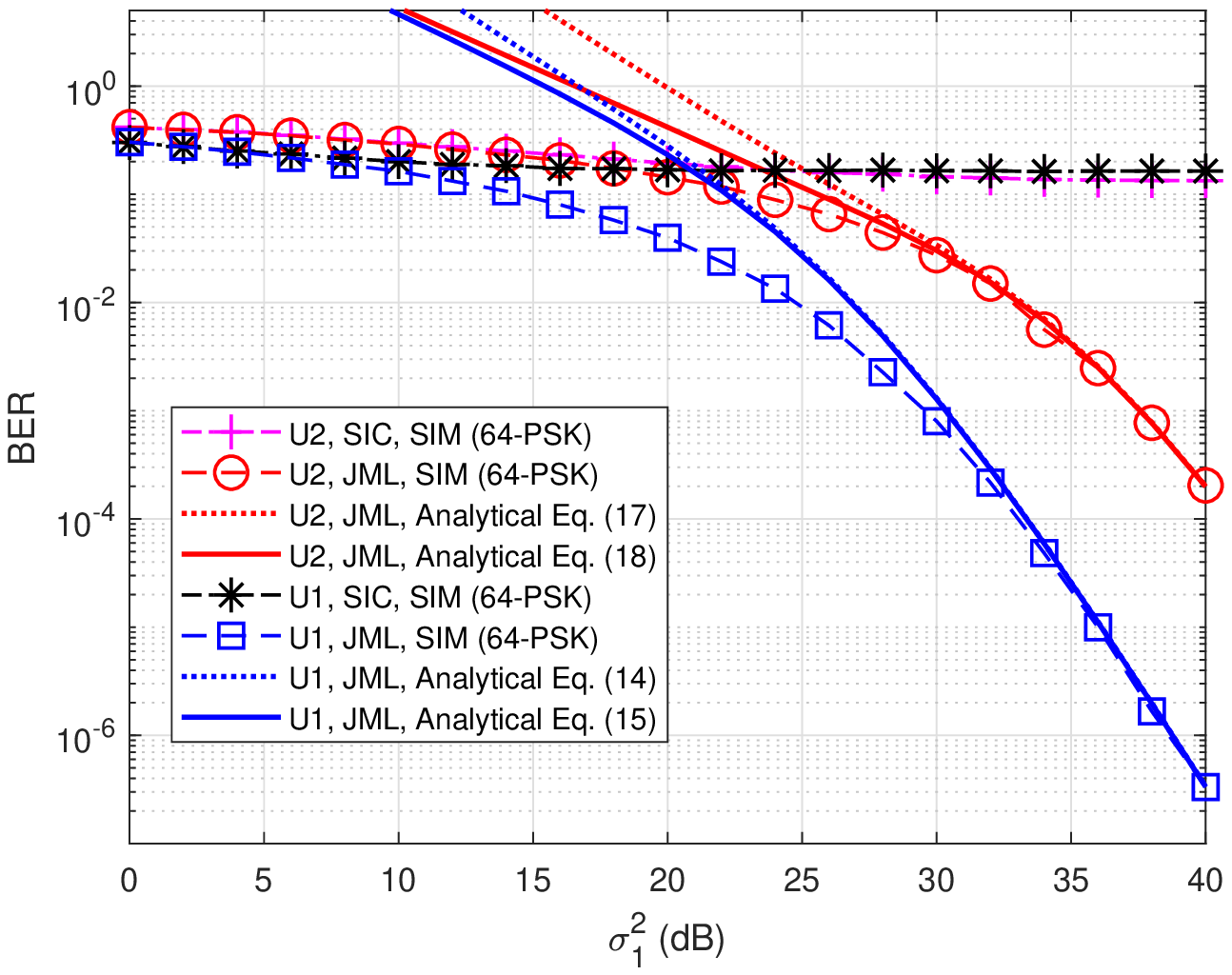}
\label{ber3}}
\caption{{Error Performance of uplink SIMO-NOMA for $M=M_1=M_2$ when $L=4$ antennas at BS: BER vs $\sigma_1^2 (dB)$.}}
\label{vs_SNR}
\end{figure*}




As a general rule, as the channel conditions aggravate, a stronger (low order) modulation is chosen, while a high order modulation is preferred for a better channel condition. As U2 is suffering from the worst channel, then we choose a lower modulation order for U2 compared to that of U1. Hence, in the second scenario, the Fig. 4 shows the BER performances for uplink SIMO-NOMA for $(M_1,M_2 )=(64,32)$ and $(M_1,M_2 )=(64,16)$, respectively. For the case of $(M_1,M_2 )=(64,32)$ we have the same pattern for the JML detector, always the performances of U1 outperform those of U2. We can clearly notice that the BER of U2 begins to approach the BER of U1 and this is due to the improvement in distance between  symbols $(M_2=\sfrac {M_1}{2})$ which compensates for the effect of the severe channel. On the other side, when we increase the difference between modulation beyond 2 bits $(M_2=\sfrac{M_1}{4})$, the performance of U2 (that uses a low number of bits to construct constellation) outperforms the U1 performance and it is due to the increase of the minimum distance between symbols in low constellations. In the adaptive modulation case, we can also observe the superiority of the MRC-JML over MRC-SIC detector. Again, one can easily observe the full diversity order for both users. This reveals the effectiveness of the JML in the uplink NOMA cases, and we believe that the JML eliminates the detection problem in the uplink NOMA and this will accelerate the uplink NOMA studies.

\begin{figure}[htbp]
\centering
\subfloat[{$M_1=64$ and $M_2=32$}]{\includegraphics[width=0.85\columnwidth]{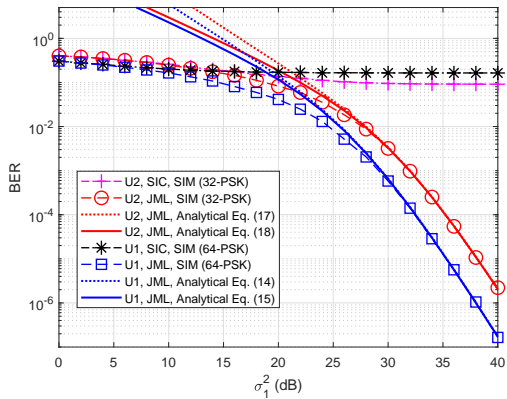}
\label{berM1}}\\
\subfloat[{$M_1=64$ and $M_2=16$}]{\includegraphics[width=0.85\columnwidth]{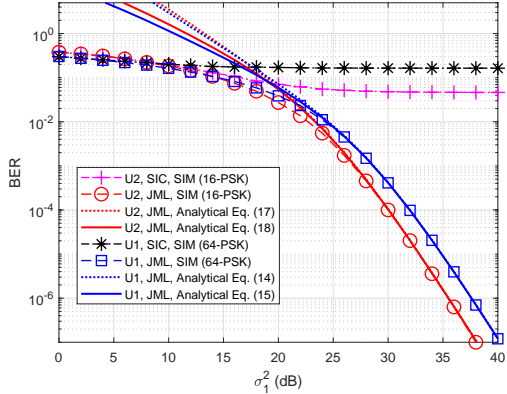}
\label{berM2}}
\caption{{Error Performance of uplink SIMO-NOMA when $L=4$ antennas at BS:  BER vs $\sigma_1^2 (dB)$.}}
\label{vs_SNR_difM}
\end{figure}



\section{Conclusions}
In this paper, we study the error performance of the uplink SIMO-NOMA system. Based on the union bound approach, we derive an upper bound for the BER in the presence of JML errors over Rayleigh fading channel. This bound was calculated for two users emitting with different M-ary PSK modulation by using an adaptive modulation technique. Based on the extensive simulations, the analytical results match well with the numerical results. The computer simulations show clearly that the MRC-JML outperforms MRC-SIC significantly and it eliminates the error floor in the uplink NOMA completely. Besides, the JML achieves the full diversity order and this proves the power of the JML rather than SIC detectors for the uplink NOMA. In this paper, we analyzed the error performance of the JML detector for arbitrary M-ary PSK and two users. We believe that our results can be used to help analyze the uplink NOMA system with different M-QAM modulations and arbitrary users. These are the future research directions.


\begin{thebibliography}{00}
\bibitem{b1} Q. C. Li, H. Niu, A. T. Papathanassiou and G. Wu, "5G Network Capacity: Key Elements and Technologies," in IEEE Veh. Technol. Mag., vol. 9, no. 1, pp. 71-78, 2014.
\bibitem{b2} S. M. R. Islam, N. Avazov, O. A. Dobre and K. Kwak, "Power-Domain Non-Orthogonal Multiple Access (NOMA) in 5G Systems: Potentials and Challenges," in  IEEE Commun. Surveys Tuts., vol. 19, no. 2, pp. 721-742, 2017.
\bibitem{b3} H. Tabassum, M. S. Ali, E. Hossain, M. J. Hossain and D. I. Kim, "Uplink Vs. Downlink NOMA in Cellular Networks: Challenges and Research Directions," 2017 IEEE 85th  IEEE Veh. Technol. conf., Sydney, NSW, Australia, 2017, pp. 1-7.
\bibitem{b4} J. Wang, B. Xia, K. Xiao, Y. Gao and S. Ma, "Outage Performance Analysis for Wireless Non-Orthogonal Multiple Access Systems," in IEEE Access, vol. 6, pp. 3611-3618, 2018.
\bibitem{b5} X. Liang, X. Gong, Y. Wu, D. W. K. Ng and T. Hong, "Analysis of Outage Probabilities for Cooperative NOMA Users with Imperfect CSI," 2018 IEEE 4th Info. Technol. and M.E. Conf. (ITOEC), Chongqing, China, 2018, pp. 1617-1623.
\bibitem{b6} A. Agarwal, R. Chaurasiya, S. Rai and A. K. Jagannatham, "Outage Probability Analysis for NOMA Downlink and Uplink Communication Systems With Generalized Fading Channels," in IEEE Access, vol. 8, pp. 220461-220481, 2020.
\bibitem{b7} F. Kara  and H. Kaya, “BER performances of downlink and uplink NOMA in the presence of SIC errors over fading channels,” IET Commun., vol. 12, no 15, pp. 1834-1844, 2018.
\bibitem{b8} F. Kara and H. Kaya, "Error Probability Analysis of NOMA-Based Diamond Relaying Network," in IEEE Trans. Veh. Technol., vol. 69, no. 2, pp. 2280-2285, 2020.
\bibitem{b9} J. S. Yeom, H. S. Jang, K. S. Ko and B. C. Jung, "BER Performance of Uplink NOMA With Joint Maximum-Likelihood Detector," in IEEE Trans. Veh. Technol., vol. 68, no. 10, pp. 10295-10300, 2019.
\bibitem{b10} F. Kara and H. Kaya, "Improved Error Performance in NOMA-based Diamond Relaying," 2020 IEEE  Microw. Theory Techn. in Wireless Commun. (MTTW), Riga, Latvia, 2020, pp. 151-156.
\bibitem{b11} M. B. Shahab, S. J. Johnson, M. Shirvanimoghaddam, M. Chafii, E. Basar and M. Dohler, "Index Modulation Aided Uplink NOMA for Massive Machine Type Communications," in IEEE Wireless Commun. Lett., vol. 9, no. 12, pp. 2159-2162, 2020.
\bibitem{b12}  3GPP, “NR; Physical Channels and Modulation,” Mar. 2021, TS 38.211, Release 16, v16.5.0.
\bibitem{b13} H. Carvajal, N. Orozco and C. de Almeida, "On the Performance of Maximal Ratio Combining and Maximum Likelihood Detection in M-QAM/MC-CDMA Systems," 2014 Int. Telecom. SYM. (ITS), Sao Paulo, Brazil, 2014, pp. 1-5.
\bibitem{b14} D. Tse and P. Viswanath, Fundamentals of Wireless Communication. Cambridge University Press, 2005.
\bibitem{b15} Jianhua Lu, K. B. Letaief, J. C. -. Chuang and M. L. Liou, "M-PSK and M-QAM BER Computation Using Signal-Space Concepts," in IEEE trans. commun., vol. 47, no. 2, pp. 181-184, Feb. 1999.
\bibitem{b16} William C. Y. Lee , "Mobile Communications Engineering: Theory and Applications," Second Edition, The McGraw-Hill Companies, Inc. 1998.

\end{thebibliography}
\end{document}